\documentclass[sn-mathphys]{sn-jnl}

\usepackage{mdframed}
\usepackage{xcolor}

\theoremstyle{thmstyleone}%
\theoremstyle{thmstyletwo}%

\theoremstyle{thmstylethree}%

\usepackage{verbatim}

\begin{document}

\title[Future Directions in Human Mobility Science]{Future Directions in Human Mobility Science}

\author[1,7]{\fnm{Luca} \sur{Pappalardo}}

\author[2,3]{\fnm{Ed} \sur{Manley}}

\author[4,5]{\fnm{Vedran} \sur{Sekara}}

\author*[6]{\fnm{Laura} \sur{Alessandretti}}\email{lauale@dtu.dk}

\affil[1]{\orgdiv{Institute of Information Science
and Technologies}, \orgname{National
Research Council (ISTI-CNR)}, \orgaddress{\city{Pisa}, \country{Italy}}}

\affil[2]{\orgdiv{School of Geography}, \orgname{University of Leeds}, \orgaddress{\city{Leeds}, \country{UK}}}

\affil[3]{\orgdiv{Leeds Institute for Data Analytics}, \orgname{University of Leed}, \orgaddress{\city{Leeds}, \country{UK}}}

\affil[4]{\orgname{IT University of Copenhagen},  \orgaddress{\city{Copenhagen},  \postcode{2300}, \country{Denmark}}}

\affil[5]{\orgname{UNICEF},  \orgaddress{\city{New York},  \postcode{10017}, \country{USA}}}

\affil[6]{\orgdiv{DTU Compute}, \orgname{Technical University of Denmark}, \orgaddress{\city{Kgs. Lyngby}, \postcode{2800}, \country{Denmark}}}

\affil[7]{\orgname{Scuola Normale Superiore}, \orgaddress{\city{Pisa}, \country{Italy}}}


\abstract{We provide a brief review of human mobility science and present three key areas where we expect to see substantial advancements. We start from the \emph{mind} and discuss the need to better understand how spatial cognition shapes mobility patterns. We then move to \emph{societies} and argue the importance of better understanding new forms of transportation. We conclude by discussing how \emph{algorithms} shape mobility behaviour and provide useful tools for modellers. Finally, we discuss how progress in these research directions may help us address some of the challenges our society faces today.}




\maketitle

\section*{Introduction}\label{sec1}

\begin{figure}[]
\centering
\includegraphics[width=0.99\textwidth]{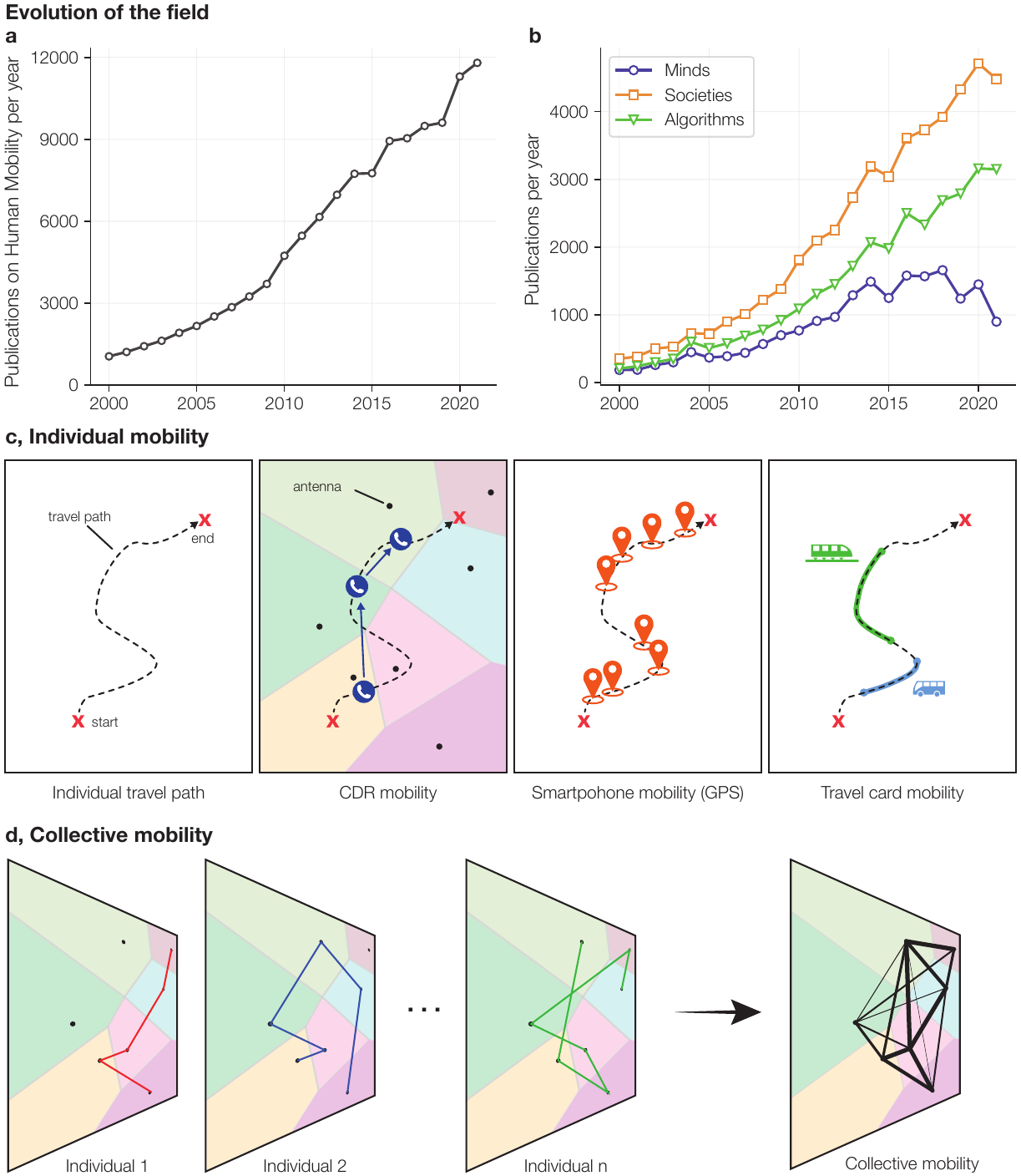}
\caption{\textbf{Human Mobility data sources}.\textbf{a)} The yearly number of publications on Human Mobility as obtained by querying Google Scholar \cite{noruzi2005google} with $Q =$ ("human mobility" OR "mobility network" OR "mobility flow" OR "flow of mobility" OR "mobility dynamics" OR "mobility patterns").
\textbf{b)} The number of Human Mobility publications for Minds, Societies and Algorithms. Results are obtained by querying Google Scholar for Minds (circles): $Q$ AND "spatial cognition" OR "psychology" OR "navigation" OR "cognitive mechanism" OR "spatial memory"); Societies (squares): $Q$ AND ("transportation" OR "public transport" OR "multimodal" OR "sustainable"); Algorithms (triangles): $Q$ AND ("machine learning" OR "artificial intelligence" OR "AI" OR "computational"). 
\textbf{c)} Individual mobility paths. From left to right: mobility from CDR data, GPS data collected via smartphones apps, and travel paths estimated from smart-cards. For CDR data, mobility is re-constructed as trips between cell-towers (black dots) whenever individuals issue/receive an sms or call (the background shapes describe the corresponding Voronoi tessellation). GPS data positions are recorded with high spatial resolution, typically at fixed time intervals or whenever the accelerometer of the mobile device registers a change. Travel card data capture portions of a trips that are traveled using public transportation.
\textbf{d)} Collective mobility data is obtained through aggregating individual-level data. Example show the aggregation of CDR mobility from multiple individuals.} \label{fig1}
\end{figure}

It was around 130 years ago when scientists first found evidence for universal properties in human mobility patterns. 
Looking at the places of birth and residence of individuals living in the UK, German-British geographer Ernest George Ravenstein found out that simple laws could capture migration within and across the country \cite{ravenstein1889laws, ravenstein1876birthplaces}. 
It was a pioneering result that laid the ground for what is still -- over a century later -- a very active research field \cite{barbosa2018human, Wang2019-io}.
These achievements are even more remarkable in view of the limitations of the census registry data used at the time – coarse spatial resolution and no time information – and the titanic work carried out by statisticians and clerks who would manually process handwritten records into statistical tables. 
Since then, the field has evolved dramatically (see Figure \ref{fig1}a and b). From migration to navigation, via commuting and running day-to-day errands, we have now uncovered regularities in many human activities that involve movements \cite{barbosa2018human, Wang2019-io}. 

On the one hand, these developments were made possible by the availability of increasingly detailed and comprehensive data. 
In the last 20 years, the field of Human Mobility has evolved hand-in-hand with the development and now widespread diffusion of positioning technologies, which were first embedded in mobile phones and vehicle navigation systems, and are now in smartphones and wearable devices (see Figure \ref{fig1}c).
Today, it is estimated that 6.6 billion individuals own a smartphone, meaning that about 85\% of the world population move with a location sensor in their pockets \cite{o2021smartphone}.  

Concurrently, the study of behavioural data with expanding scale and complexity has been empowered by tremendous advances in computation. Cloud and distributed computing have facilitated the storage and analysis of massive spatio-temporal datasets; computational modelling through simulations has helped study complex, nonlinear systems, previously considered intractable \cite{Alessandretti2020,Barbosa2015-sf,bongiorno2021vector,ijgi10090599,Pappalardo2015,schlapfer2021universal}; statistical learning algorithms have been used to extract patterns from complex behavioural data \cite{alessandretti2017multi,kraemer2020mapping,Noulas2012-ku}, predict \cite{Burbey2012-tx,Calabrese2010-el} and generate \cite{dai2015personalized,Luca2021-dv,Pappalardo2018,Simini2021} mobility trajectories with unprecedented success. 

From a statistical standpoint, a trajectory is understood as an alternation of stays -- when an individual spends time at a given location -- and displacements, or the navigation between locations  \cite{zheng2015trajectory, zheng2014urban, Luca2021-dv, barbosa2018human}. 
Considering the sequences of stays alone, we have achieved an understanding that was simply unimaginable $20$ years ago.
We have been able to capture the statistics of stays across an almost exhaustive range of spatial and temporal scales, from short visits to shops up to residential moves \cite{alessandretti2017multi, Wang2019-io, pappalardo2021evaluation, fudolig2021internal, moro2021mobility, pappalardo2016analytical}.
We have been able to describe how trip purpose \cite{xue2022leveraging,aleta2022quantifying,lucchini2021living} and urban characteristics \cite{borst2009influence,coutrot2022entropy} impact travel by enriching sequences of stays with data describing the built and natural environment.
Further, it has been possible to describe differences across individuals, based on, for instance, gender, age, socio-economic status, and country of residence \cite{kraemer2020mapping, gauvin2020gender, macedo2022differences, barbosa2021uncovering, moro2021mobility}.

Remarkably, notwithstanding the individuality of each person, we have discovered that many properties of travel patterns are shared across people. 
This includes, for example, the probability to travel at a given distance \cite{alessandretti2017multi, gonzalez2008understanding,gallotti2016stochastic, pappalardo2013understanding}; the frequency at which we visit \cite{schlapfer2021universal, gonzalez2008understanding,song2010limits}, explore \cite{Pappalardo2015, scherrer2018travelers} and renew location \cite{Alessandretti2018-aq, barbosa2015effect}; the way we allocate time among places \cite{Song2010} the predictability of human mobility \cite{song2010limits, lu2013approaching, smith2014refined, cuttone2018understanding, ikanovic2017alternative}; the hierarchical nature of travel \cite{Alessandretti2020}; the spatio-temporal structure of individual mobility networks \cite{schneider2013unravelling, Yan2017}; and the link between individual and collective flows \cite{schlapfer2021universal, Simini2012-gq, Yan2017} (see Figure \ref{fig1}d).

Why these regularities emerge, however, is not yet entirely understood. 
Agent-based modeling have revealed how the observed patterns could result from simple decision-making mechanisms that are shared across individuals \cite{Song2010,Alessandretti2020,schlapfer2021universal,Pappalardo2015,gallotti2016stochastic}, but, in many cases, these hypotheses lack empirical demonstration.

As we see more advances in computation -- from deep generative models to federated learning -- and embrace new technologies -- from mobile electroencephalography (EEG) to self-driving cars and sensors deployed in smart cities -- the study of human mobility will continue to flourish. 
In this Perspective, we discuss three key areas where we expect to see substantial advancements. 
We will start precisely from the \emph{mind} and discuss the need to better understand how cognition shape human movements, especially in navigation. 
We will then move to \emph{societies} and discuss the critical need to better understand new forms of transportation and the opportunity of studying social mixing in cities. 
We will conclude with a focus on \emph{algorithms}, and Artificial Intelligence (AI) in particular, discussing how they are not only helping us model human mobility but are also impacting mobility behaviour.
Noting the critical role played by empirical data across all areas of Human Mobility Science, in Box 1 we introduce some critical issues and potential solutions surrounding the availability and quality of data sources. 
 \ \\
\begin{mdframed}[backgroundcolor=gray!20] 
\small{
\textbf{Box 1: Issues with mobility data} \\ 
The majority of datasets that are used to estimate human mobility (from mobile phone records, smartphone apps' records, social media check-ins, and travel card data) are originally not designed for this purpose.
Therefore, different biases and patterns are introduced into the mobility traces, depending on which technology was used to collect the data.
For instance, mobility inferred from Call Detail Records (CDR) or Data Records (XDR), which are passively collected by mobile network operators for billing purposes, can suffer from multiple biases~\cite{sekara2019mobile, schlosser2021biases, zhao2016understanding}.
For individuals to be included in the mobility sample, they must own a mobile phone, and mobile phone ownership has been shown to be biased towards predominantly wealthier, male, younger, and better educated populations~\cite{blumenstock2010mobile,wesolowski2012heterogeneous,wesolowski2013impact}.
It is worth noting, however, that the severity of the issue depends on the region of study. 
Statistics from 2020 show that there are large variations between countries; in high-income countries there are 122 mobile phone subscriptions per 100 people, while in low-income countries the number is only 59~\cite{ITU-stats}.
Similarly, to be included in an CDR or XDR dataset requires individuals to actively make or receive calls and texts, or use mobile data.
While pre-paid mobile phone subscriptions in a lot of countries include generous limits on calls, texts and data usage, for lower income countries, pay-as-you-go subscriptions are still popular.
For datasets collected in these places, it is vital to be aware that poorer populations will limit their mobile activity as making a call, sending a text message, or using mobile data can be very costly.
As a result, they will generate less activity data, and their mobility patterns will be less complete~\cite{leo2016socioeconomic,schlosser2021biases}.
In addition, due to commercial interests, not all of the regions are equally covered by mobile phone operators \cite{ricciato2015estimating}, which effectively limits the resolution with which mobility traces can be inferred.
Mobility traces in rural areas are sparse as large areas (100km$^2$ or larger) are often covered by only one antenna, while, for urban areas, mobility traces are more fine-grained as the density of antenna is much higher (down to 1 per 100m$^2$).
Understanding which populations are included in the data, at which resolutions their traces can be reconstructed, and how the data is generated, are vital first steps in working with biased and incomplete data.

Similar considerations are needed for GPS-inferred mobility data from smartphone apps and social media platforms~\cite{acosta2020quantifying,barbosa2021uncovering}.
Datasets that are based on apps are often infused by the goals of the platform, the app, and its owners, which ultimately introduces various patterns into the data. For instance, apps that collect GPS data may be socially engineering and nudging their users to get them to use the app more often.
This can happen algorithmically or through some predefined heuristics. Nonetheless, this is an invisible occurrence, which is difficult to detect, but that will inevitably seep into the mobility data~\cite{salganik2019bit}.

Biases in data, issues regarding mis- or non-representation, resolution differences, and data skews will manifest differently depending on the data source at hand, and on the processes and methods used to collect it.
Unless these issues are addressed and accounted for, they might propagate and affect any insights derived from the data~\cite{schlosser2021biases,acosta2020quantifying}.

Unfortunately, there are no easy or standardized fixes.
The literature is still sparse on how to properly address these issues~\cite{tizzoni2014use,pestre2020abcde}, compounded by the lack of ground-truth estimates of mobility.
Certain techniques are useful for alleviating some of these issues.
Examples of these techniques include sampling, post stratification, inclusion of synthetic data from null models, such as the gravity or radiation models, and a combination of travel patterns from multiple data sources.

In addition, mobility datasets are highly sensitive and can contain detailed information regarding people’s whereabouts, financial standing~\cite{moro2021mobility}, social-graphs~\cite{blondel2015survey}, and countless other behavioral patterns~\cite{de2018privacy}. 
Attitudes towards individual data collection vary widely, with some people fearful of privacy violations, others uncritically enthusiastic about technology, and some resigned to the practice \cite{rzeszewski2018care}.
Users tend to be willing to share their data if they anticipate substantial benefits, even when they have privacy concerns \cite{gerber2018explaining}.
Importantly, these diverse attitudes can further introduce biases into spatial datasets.
The difficulties around obtaining informed consent for use of these data have no clear solution~\cite{taylor2016no}.
Currently, there is no agreed-upon standards, but it is of utmost importance that this data is handled in safe, privacy-preserving, and ethical ways~\cite{maxmen2019surveillance}.
}
\end{mdframed}

\section*{Mind: Spatial Cognition and Mobility Choices}\label{cog mechs}

Many everyday travel tasks, from planning one’s own way in space, to locating resources, rely on our ability to organise and retrieve spatial knowledge. The human mind, however, forms a limited and distorted picture of the physical space~\cite{tversky1993cognitive}. Understanding how the human brain organises spatial knowledge,
or builds so-called ``cognitive maps'', is central to understanding mobility behaviour, and ultimately to planning human-centred cities and transportation systems.

Despite all of the advancements in the field of Human Mobility, however, there is still little connection to the broader research on the cognitive mechanisms underpinning travel and navigation. The structures in our brains – evolved over thousands of years – determine how we recall directions, associations between locations, and distances between those points, and ultimately shape the choices that we all make to move through space. While we know that our perceptions and memories of space are far from perfect, we have less understanding of how these impact human travel. Three areas in the intersection of human cognition and mobility are prime for further exploration: spatial memory and representation, navigation and mobility choice strategies, and influential contexts and personal characteristics. 

\subsubsection*{Memory and Representation}
\begin{figure}[h]
\centering
\includegraphics[width=0.95\textwidth]{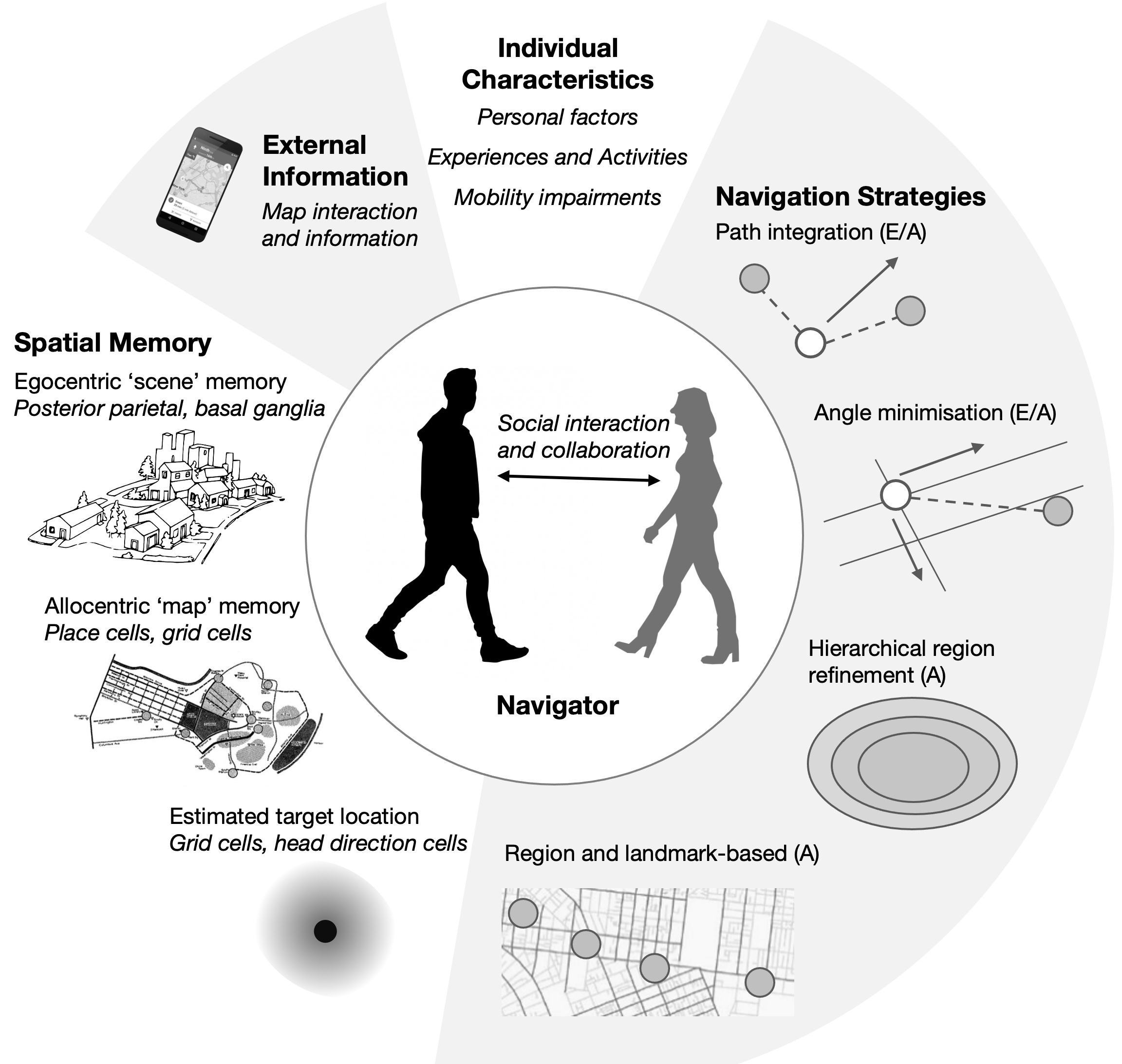}
\caption{\textbf{Summary of determinant factors and cognitive structures in spatial cognition} a) Individual and contextual factors relating to the availability of external information, social interactions and collaboration, and individual characteristics, b) Spatial memory consists of interacting representations based on features, scenes, and the association between them within map-like models. Head direction cells keep track of the location of an unseen target; c) Different strategies shape navigation, working in conjunction, and with other choice architecture, and drawing on different memory representations. These strategies are inherently subjective, potentially results in objectively sub-optimal route choices.}\label{fig2}
\end{figure}

As a person moves through space, they build a memory of the distance and association between places (see Figure \ref{fig2}). 
Initially, the memory consists of a set of observed ``features'' usually aligned along a route (\emph{egocentric} representation, meaning, relative to the self). 
As one's experience of a place increases, these ``features'' are integrated into a structure that captures their relationship in space (\emph{allocentric} representation) independent of the location of the individual. 
The resulting representation of spatial memory has been called a ``cognitive map''~\cite{tversky1993cognitive}, although the co-presence of both representations means that a single map is difficult to isolate. 

Those memorable ``features'' in geographic space play differing roles in shaping memory. 
Landmarks – either culturally significant or merely visually salient – are noted as important features in encoding spatial memory. 
Major road intersections~\cite{Sadalla1980}, routes~\cite{weisberg2016some} edges of regions and zones~\cite{Newcombe1982}, are other key features in spatial memory. 
Neuroscience would appear to validate many of these proposals. 
Within the hippocampus region of the brain, a set of cells encode the position of common objects, preserving an individual map-like representation of space: place cells encode salient locations, border cells indicate regional boundaries, and grid cells capture distance between spaces (see Moser et al. \cite{moser2008place} for a review). 
These mechanisms are supported by head direction cells and goal direction cells, that describe subjective position. 
Findings that features are organized within hierarchies \cite{hirtle_evidence_1985}, in which objects (for instance, neighbourhoods and streets) are organised within nested structures, has also been recently established in cognitive planning processes~\cite{balaguer2016neural}.

It is upon these cognitive representations that mobility choices are made.
Yet despite the prevalence of computational models of mobility behaviour, there have been few linkages made between spatial cognition and spatial data representation. 
An important, qualitative framework in instigating research around this area is Kevin Lynch’s Image of the City \cite{lynch_image_1960}, which proposes that perceptions of cities can be deconstructed into five elements: edges, nodes, landmarks, regions, and paths. 
The Image of the City has been highly influential, and led to recent implementations in modern spatial data sources \cite{filomena_computational_2019}. 
Focusing on specific features of urban space, space syntax researchers have proposed models that capture saliency within road networks, and others have focused on recognition of prominent landmarks \cite{winter2008landmark}. Researchers in robotics, on the other hand, have sought to explicitly replicate human spatial learning \cite{kuipers_modeling_1978, chown_prototypes_1995}. 
However, aside from prototypical examples \cite{manley2021spatial}, there is an absence of comprehensive proposals that capture how real-world geographic space is learnt, stored and recalled, and how knowledge varies across a population. Such a model (essentially a ``cognitive GIS'') would provide a novel basis for measuring how humans and space interact. 

\subsubsection*{Navigation and Mobility Choices}
The role of spatial cognition arises in many facets of human mobility, including choosing destinations, transport modes, and routes, or navigating in new environments.
While the majority of prior studies focus on navigation, these mechanisms influence wider travel decisions.

The findings from cognitive science put in doubt persistent assumptions in urban and transportation science relating to utility-optimising behaviour when choosing a route to destination. 
Traditionally, route choice has been modelled as a logistic choice between a set of alternative routes, calibrated using data collected through interviews and surveys \cite{ben1984modelling}, which suffer of small sample sizes, missing data, and self-reporting biases. 
These alternatives are typically modelled as shortest paths (for instance, paths that take the least time or distance), but recent findings in real-world settings have established that pedestrians and vehicle drivers do not tend to follow optimal routes, and are instead influenced by urban features and goal distance and direction \cite{manley_shortest_2015, malleson_characteristics_2018, bongiorno2021vector, lima2016understanding, sevtsuk_role_2022}. 
The structure of the street network clearly also plays a role \cite{coutrot2022entropy}, as may greenery \cite{miranda2021desirable}, traffic volume \cite{borst2009influence}, and noise \cite{guo2013pedestrian}. 
Yet, there are considerable opportunities to strengthen and expand these findings and develop more sophisticated models, informed by our knowledge of the underlying cognitive mechanisms. 

Further opportunities arise when considering the mechanism by which decisions are modelled, and how these theory are implemented into predictive models. 
Aside from discrete choice, approaches drawn from behavioural economics, cognitive modelling \cite{urban2000pecs,georgeff1999belief}, and reinforcement learning have been considered in shaping mobility choices.
The influence of bounded rationality is increasingly recognised as fundamental to choice \cite{bongiorno2021vector}, and heuristic frameworks have been proposed \cite{manley2015heuristic}. 
More recently, opportunities that blend theoretical and machine learning methods have emerged, including from those in traditional methods \cite{aboutaleb_discrete_2021} and artificial intelligence \cite{banino_vector-based_2018, mirowski_learning_2018}. 

Despite our improving understanding of spatial navigation, fewer linkages between spatial cognition and more 'strategic' travel choices (e.g. destination choice, travel mode, departure time) have been established. 
Recent promising research has begun to consider how physiological \cite{hancock2023utilising} and fMRI data \cite{qin2023perceptions} can built into models of spatial and temporal perception and mobility choice.
While these findings contribute towards how we model spatial and temporal perception, they must be integrated within a set of behavioural determinants, including the latest, complementary theories of travel choice.

\subsubsection*{Characteristics, Context, and Information}
Beyond the mechanisms influencing mobility, a range of additional factors have been shown to mediate mobility. 
Personal characteristics are particularly important. 
For instance, navigation efficiency is lower in females and older people \cite{coutrot_global_2018}. 
The location of your upbringing also plays a role  \cite{coutrot2022entropy}. 
Mobility or visually impaired travellers face additional limits on mobility, for a large part due to the environment. 
The context in which the traveller is will be important too. 
For instance, a tourist unfamiliar with an environment navigates differently when compared to a local, but local knowledge varies considerably too, based on experience \cite{montello1998new}. In addition, the presence of friends or family introduces a social element to navigation, necessitating cooperation and information sharing \cite{mavros_collaborative_2022}.
Traffic and congestion also prompt en-route changes to plans \cite{spiers2008dynamic}. Finally, the role of navigation support devices (for instance, satellite navigation devices and maps) influences both behaviour and spatial learning \cite{ishikawa2008wayfinding, gardony2013navigational}.

While the role of cognitive maps is co-dependent with context and characteristics, many of the existing studies have been conducted in small-scale, controlled settings, and we lack a systematic, intersectional understanding of cognitive factors impacting human movement.
Collection of data integrating these mobility behaviour with determinant factors is challenging at the larger scale, facing many of the same issues associated with other mobility studies (see Box 1).
Computational analyses of passively collected data for large populations, especially if combined with experimental data, may be key to pinpoint some of the cognitive mechanisms underlying travel in real-world settings. 

Ultimately, understanding the fundamental mechanisms underlying travel choices, human navigation and routing will bridge the literature streams of spatial cognition, human mobility and transportation engineering and will make possible to generate more realistic traffic and transit models.

\begin{figure}[htb!]
    \centering
    \includegraphics[width=\linewidth]{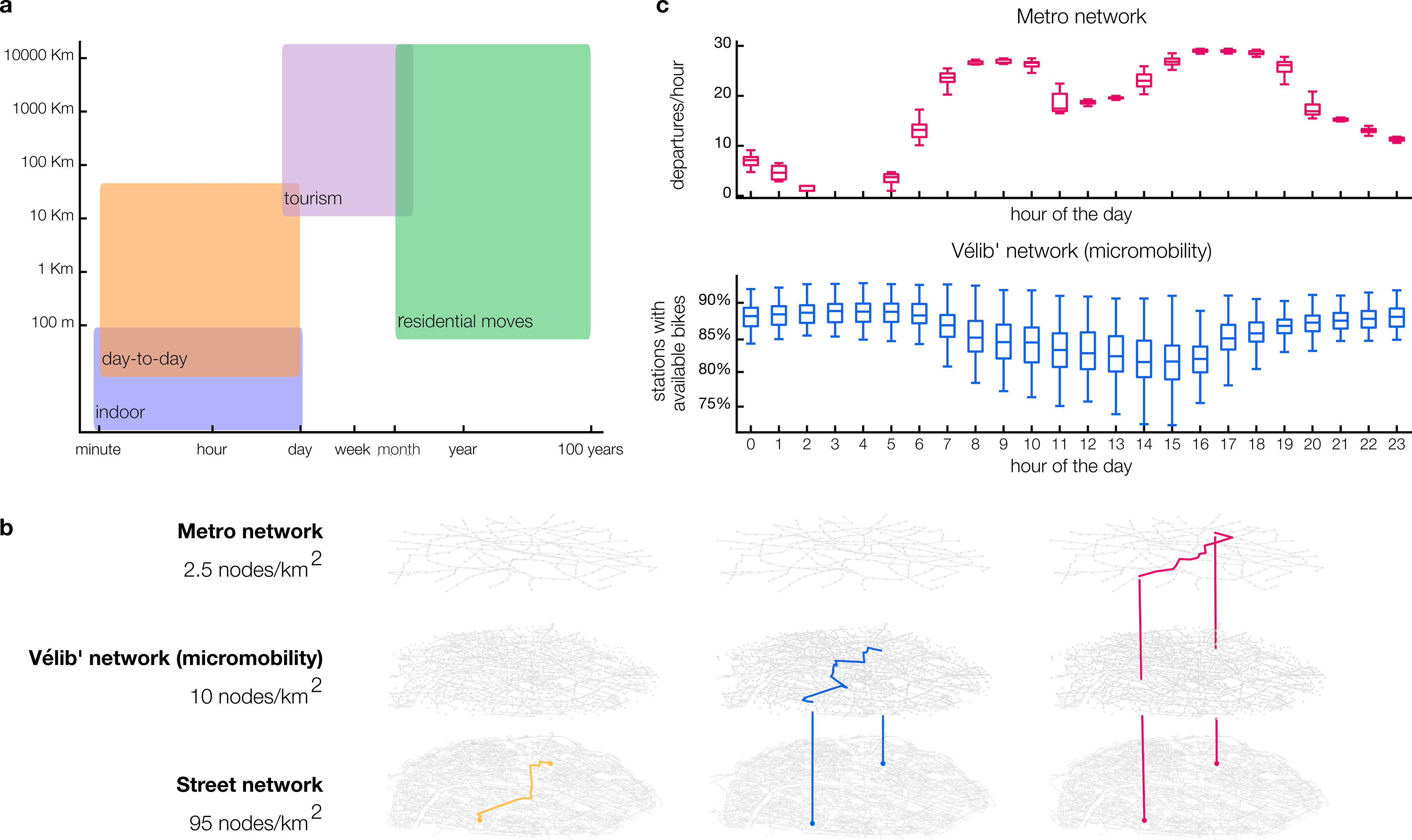}
    \caption{\textbf{The complex spatio-temporal nature of multimodal transport systems.} \textbf{a)} Human mobility occurs over a large range of spatial and temporal scales. Urban mobility deals mostly with day-to-day displacements, through different transport modalities. \textbf{b-c)} Example of multimodal transportation network in the city of Paris. In panel \textbf{b} the bottom layer corresponds to the walking layer (street network); the mid layer to the Velib network (micromobility bikes), where links are cycle lanes, and nodes are bike docks; the top layer is the metro network. Layers have widely different spatial densities (reported in the figure). While the walking network is static, such that travelling from a given origin to a given destination is always possible (left), travelling along the micromobility layer is subject to the availability of bikes (center) and trips along the metro network are subject to the public transport schedule (right). \textbf{c)} Boxplots displaying the availability of transport options in the Paris transport network as a function of the time of the day: (top) number of vehicle departures/hour across Paris metro stations; (bottom) fraction of Velib stations with at least one bike. For each hour of the day: the horizontal line corresponds to the median value, the box contains 50\% of the data, whiskers contain 90\% of the data. Street network data from OpenStreetMap \cite{OpenStreetMap}; Metro data from RATP \cite{ratp}; Vélib' data from data.gouv.fr \cite{datagouvfr}}  
    \label{fig:multimodality}
\end{figure}

\section*{Societies: Cities and Transportation systems}
Many global cities are striving to shift from a car-centric approach to a solution that integrates several modes -- including walking, cycling, and public transport~\cite{alessandretti2022multimodal} -- to support urban day-to-day trips across a range of spatial and temporal scales (see Figure \ref{fig:multimodality}a). 
The scientific understanding of human mobility is key to help with this transition, because it provides policy-makers with tools to predict future states of urban transport demand and assess the impact of new policies.
There are critical gaps, however, in our understanding of multimodal urban travel. 

A first important challenge concerns the limited availability of data capturing multimodal trips.
For example, mobility patterns derived from travel cards describe the use of public transportation, but do not provide the full picture on how people travel from the origin to the destination of a trip \cite{mohamed2016clustering,zhong2015measuring,zhong2016variability}. 
Additionally, most of the existing data originates from developed countries, and the complexity of urban transportation in the global south, where the largest and fastest-growing cities are found \cite{nations2014world}, remains largely uncharted territory (as indicated in Box 1).
Thus, it is still unclear how various transport modes and their combination meet different individual travel needs. 
A promising avenue for data collection comes from initiatives that are developing smartphone based ticketing systems for connecting passengers to multiple operators of public and micro-mobility transport \cite{bueno2021identifying,wirtz2019smartphone,baldauf2020pervasive,nahmias2018enriching}.
By gathering both GPS and transport data through multiple modes, these mobile apps have the potential to drive novel, unprecedented understanding of mobility behavior.

Trajectory data that captures transport mode choices collected for large enough populations could further help quantify how individuals are exposed to each other while using transportation. 
Understanding these urban encounters, in turn, is critical to quantify urban segregation \cite{moro2021mobility, amini2014impact} and to model epidemic spreading \cite{aleta2022quantifying}. 
Since the COVID-19 pandemic, large GPS trajectory datasets have emerged as a promising approach to quantify urban encounters, with studies reporting an unparalleled coverage of 22\% of the population in certain areas \cite{crawford2022impact}.
Future research shall address some of the challenges related to inferring contacts from mobility traces, including limits of the data in indoor settings, lack of data capturing encounters in public transportation, and sampling biases (see more details in Box 1).

A second issue surrounding the study of urban mobility is the lack of suitable models that can holistically integrate different forms of transportation.
In the last 20 years, novel shared transport solutions have become commonplace in many parts of the world. 
Micromobility (bike and scooter sharing), rides on demand (Uber and Lyft), and vehicle pooling \cite{machado2018overview} are fostering the shift from private transport to public ones. Traditional approaches have considered travel as a series of distinct processes, modelled at the aggregate scale \cite{mcnally2007four}, neglecting in dependencies between travel mode and route, for instance.
While newer approaches, such as activity-based and agent-based transport models \cite{w2016multi} provide more individual focus, aspects of choice, including routes and congestion effects, are often simplified for computational ease.

The framework of multilayer networks offers a promising and versatile tool for modeling multimodal infrastructure \cite{alessandretti2022multimodal}. Multilayer networks are organized in layers corresponding to different transport modes, where in each layer, transport infrastructure (for instance, streets, bus and subway lines, bike lanes) constitutes the links of a network, and intersections (for instance, bus stops, subway stations) constitute the nodes (see Figure \ref{fig:multimodality}b).
There is still a need, however, to comprehensively integrate the widely diverse time-dependent nature of the infrastructure characterizing each travel mode.
Public transport obeys fixed schedules, implying network links in the transit layer are only active intermittently (see an example in Figure \ref{fig:multimodality}c, top).
Private modes, instead, can be understood as static networks where links are continuously active.
Shared mobility systems, such as micro-mobility, offer different levels of flexibility, and thus lie in a broad spectrum in-between the private and public modes. 
For example, in station-based bike-share systems, transport links between any pair of stations can activate at any point in time, subject to the availability of vehicles at the departure station (see Figure \ref{fig:multimodality}c, bottom).
In car pooling, instead, links are only active when rides covering specific origin-destination routes are offered, thus requiring more planning.
Efforts to design holistic models will be crucial not only to design transport infrastructures, but also to develop comprehensive route planning algorithms that consider all available transportation modes in a truly time-dependent network.

Finally, it is important to note that, in the future, emerging technologies like autonomous vehicles, vehicle-to-vehicle (V2V) communication, and vehicle-to-infrastructure (V2I) communication will drive dramatic changes to urban mobility \cite{lu2014connected}. 

The integration of V2V and V2I technologies holds great promise in shaping the future of human mobility. 
The ability to collect real-time data on various aspects of the urban infrastructure, vehicles, and human behavior, offers a new window into the complexities of human interaction with their urban environment. This information, when analyzed and applied correctly, has the potential to revolutionize transportation systems, through enabling better traffic management, more efficient navigation, and improved road safety.
There are a range of technical challenges associated to the collection and storage of these data, including privacy concerns, interoperability issues, and regulatory hurdles, which are beyond the scope of this perspective.
In terms of computational modelling, real-time data collected for fleets of vehicles will create the need for novel approaches to mobility modeling, as it could enable to consider drivers' simultaneous choices and assign vehicles to paths in real-time in a coordinated fashion \cite{ward2016dynamic}. 

In view of a massive use of the autonomous vehicles in the near future, coordinated traffic assignment models could help dramatically reduce urban congestion. 
In this respect, it will be critical to explore methods that maintain a balance between a system optimum, meaning an assignment that is globally efficient, and a user optimum, meaning an assignment that is fair for each user \cite{morandi2021bridging}.

\begin{figure}[htb!]
    \centering
    \includegraphics[width=\linewidth]{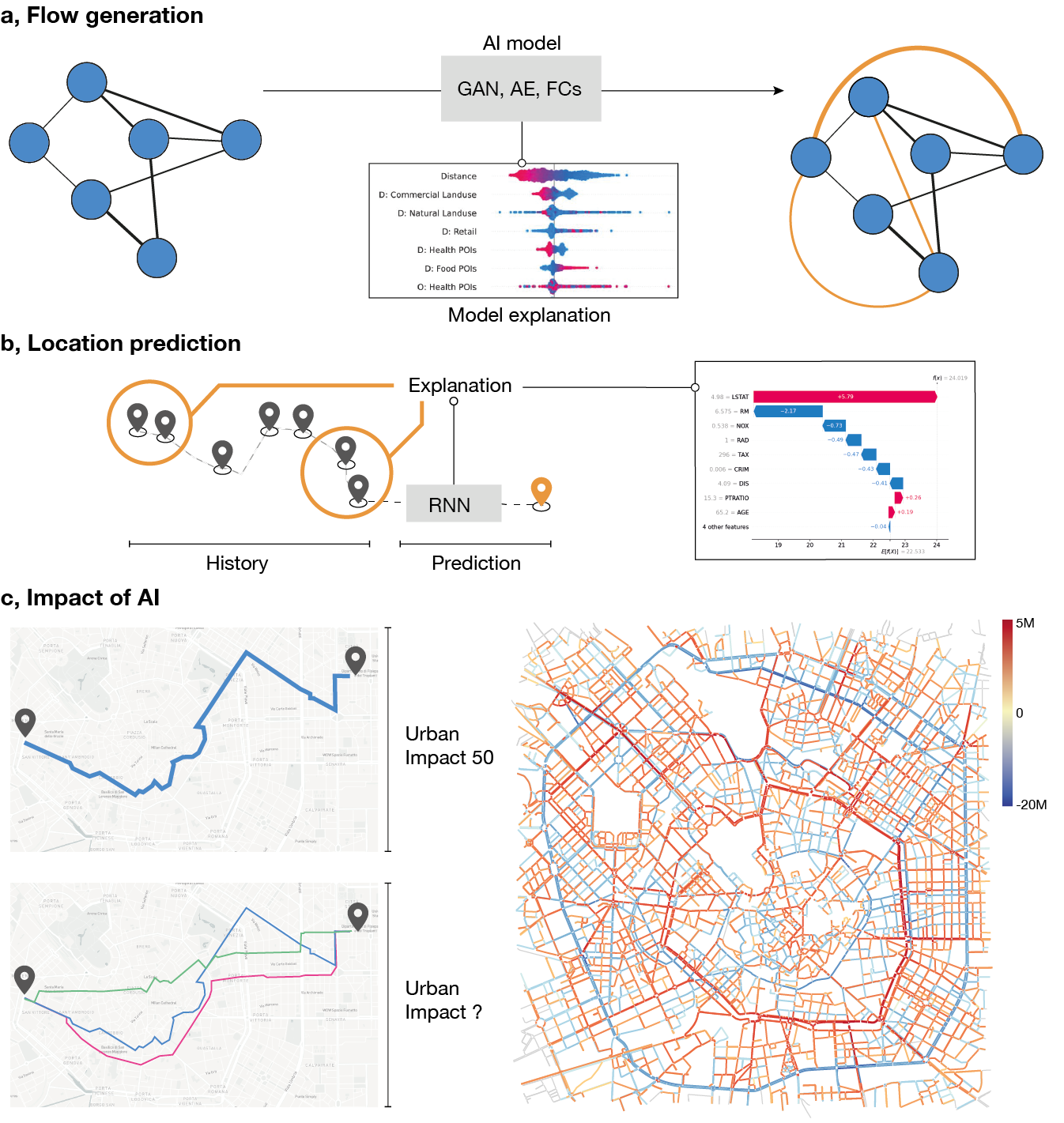}
    \caption{\textbf{Human Mobility, AI, and the Urban Environment.} (a) A schematic example of how explainable AI tools can improve a deep learning solution to a common problem in human mobility, such as flow generation. The explanation provided by the tool may indicate the importance of variables that characterize the flow's locations, for example, through a Shap-like \cite{shapely2007} explanation plot where each point represents a flow, the position on the x-axis is the Shap value, and the color indicates the feature value. Tailored explanations for human mobility are needed, and future efforts in mobility science will require defining explanations that are specifically designed for human mobility. (b) An example of how three different navigation services (NS1, NS2, NS3) may suggest different routes to vehicles with the same origin-destination pair, each path having a different impact on the urban environment in terms of externalities like CO2 emissions. Understanding the collective impact of these services and designing next-generation navigation services that can mitigate their impact while meeting user needs will be a future challenge in human mobility science.}
    \label{fig:fig4}
\end{figure}

\section*{Algorithms: computational models and AI for mobility modeling}
Motivated by human-like performance achieved in many challenging tasks in computer vision, natural language processing, medical diagnosis, and zero-sum games \cite{Sejnowski2018, LeCun2015-tu, Silver2016-tu, Titano2018-ox, Krizhevsky2017-ej, Wu2016-xa}, researchers started to use deep learning to improve solutions to fundamental mobility tasks, such as trajectory prediction and generation, and flow prediction and generation \cite{Luca2021-dv}. 
Mobility predictors and generators should capture at the same time the spatial, temporal, social, and geographic regularities that are hidden in human whereabouts, the impact of external factors, and preferences on the decision to move. 
In this regard, the use of AI in human mobility brings several advantages over traditional approaches. 

Although numerous solutions to mobility tasks have been proposed before the AI explosion \cite{Burbey2012-tx, Zheng2018-zk, Calabrese2010-el}, they typically use probabilistic or time-series-based approaches that can only partially capture human mobility dimensions. 
In particular, these traditional approaches struggle to capture complex sequential patterns in the data. 
AI approaches overtake these issues by using mechanisms such as Recurrent Neural Networks (RNNs) \cite{Rumbert1986-va} and Convolutional Neural Networks (CNNs) \cite{Krizhevsky2017-ej}. 
Moreover, AI approaches can effectively capture the influence of external factors, such as weather conditions and public events in the city, which can be efficiently handled with Fully Connected networks (FCs) \cite{LeCun2015-tu} and combined with the output of the CNN/RNN modules.

Although there is a vast literature on generative models that reproduce simple temporal, spatial, or social patterns of human mobility \cite{barbosa2018human,Pappalardo2018, Song2010, Jiang2016-bi, Wang2019-io, Barbosa2015-sf, alessandretti2022multimodal, Alessandretti2018-aq, Pappalardo2015, Toole2015-hd, zipf1946p, Simini2012-gq, Noulas2012-ku, Yan2017, Liu2020-km}, the effectiveness of these models is limited because of the simplicity of the implemented mechanisms. 
Most mechanistic generators assume that an individual relies on a simple, dichotomous decision (typically, either returning to known locations or exploring new ones \cite{Song2010,Alessandretti2018-aq}) or obeys specific laws governing human displacements (for instance, the gravity law, the radiation law, and their subsequent extensions \cite{zipf1946p, Simini2012-gq, Wang2019-io, Simini2021}). 
Recent advancements in AI, such as Generative Adversarial Networks (GANs) \cite{Goodfellow_undated-jb} and Variational AutoEncoders (VAEs) \cite{Kingma2013-eo} rely on deep learning modules to learn the distribution of data and generate trajectories or flows coming from the same distributions \cite{mauro2022generating, Luca2021-dv, bao2022covid, feng2020learning, huang2019variational, ouyang2018non}. 
GANs and VAEs are versatile and can capture different aspects simultaneously (spatial, temporal, and social dimensions) and non-linear relationships in the data that traditional approaches may fail to capture. 

Researchers are now investigating the application of Transformer networks \cite{amatriain2023transformer}, such as GPT-3 and BERT, for generating, predicting, and completing individual trajectories. 
As the impressive capacity of chatGPT\footnote{https://openai.com/blog/chatgpt/} shows, these networks have demonstrated superior performance compared to Recurrent Neural Networks (RNNs) in the area of language modeling, leading to the hypothesis that they may also be more effective in modeling human mobility. 
Recent studies provide empirical support for this hypothesis \cite{mizuno2022generation, xue2022leveraging, ma2022human, Luca2021-dv}, but the full potential of these language models is a promising research question that remains to be explored.

However, not all that glitters is gold. 
AI models (particularly those based on deep learning) are opaque \cite{Guidotti2018-de}, meaning that they are \emph{de-facto} black boxes from which it is hard to reconstruct the reasoning that led to a prediction or the generation of a trajectory/flow. 
For example, while mobility AI models rely on many features, either spatio-temporal or external ones (for instance, weather data, points of interests), it is not explicit what the role of each feature is to the model’s outcome. 
When an algorithm learns patterns from data, without human input, explainability is crucial to provide openness and transparency. 
Lack of explainability may result in erosion of trust from users or ethical issues since resulting predictions ay perpetuate the biases characterizing the empirical data (see Box 1). 
Recent studies \cite{Simini2021, Naretto_F_Pellungrini_R_Monreale_A_Nardini_FM_Musolesi_M2020-dv} show how a better interpretability can be gained by adapting explainable AI techniques developed for tabular and image data to study mobility data. 
However, these approaches have been also criticized since they can yield misleading information about the relative importance of features for predictions \cite{huang2023inadequacy, kumar2020problems}.
Therefore, it is finally time to develop explanations that are specifically tailored for human mobility, providing examples and counterexamples to validate trajectories and flows from different perspectives \cite{Luca2021-dv, jonietz2022urban}. 
Designing transparent AI mobility models or explainable mobility AI techniques is essential to gain knowledge that can be useful for potential users, such as policymakers and urban planners (see Figure \ref{fig:fig4}a and b).

A promising direction is to combine the strengths of AI models with those of mechanistic models, such as the Exploration and Preferential Return (EPR) model and its improvements \cite{Song2010, Jiang2016-bi, Pappalardo2018}, which aim to replicate the underlying processes that govern human mobility. 
Mechanistic models are interpretable by design but can capture only a limited number of mobility dimensions due to their simple mechanisms. 
Despite this, these models have a high degree of geographic transferability \cite{Luca2021-dv}, i.e., and can be used to predict locations and crowd flows in different regions.
On the other hand, AI models can capture many dimensions but need more interpretability and are highly dependent on the data used for training, which can limit their geographic transferability. 
The potential of combining these two approaches into a hybrid model is yet to be fully explored and can lead to discoveries in human mobility patterns.

\subsection*{Impact of algorithms on human mobility}
There is an increasing effort by researchers in using digital data sources to study the impact of AI mobility apps, such as new public mobility services (for instance, car-sharing, bike-sharing, ride-hailing, product and food delivery) and GPS navigation apps (for instance, Google Maps and TomTom) on several dimensions of urban welfare such as safety, air pollution, and spatial segregation \cite{Rolnick2022-wb, Bai2018-xm, Voukelatou2021-wf}. 

Although new public mobility services promise to increase the accessibility of goods and areas, their collective impact on the urban environment is largely unclear. 
The mobility task that allows these services to work is \emph{service demand prediction}, which aims to forecast (using AI) future user demand of public mobility services \cite{Yan2020-kd, Yan2019-aq, Yan_undated-iy}. 
Companies like Uber, Lyft, and Amazon provide these services every day all around the world as technology-based, on-demand, and affordable alternatives to traditional means. 
For example, people in neighbourhoods that large retailers do not reach may exploit Amazon’s same-day delivery service to avoid travelling further and paying more to obtain household necessities. 
Similarly, people without a private car may exploit ride-hailing services such as Uber to reach areas that are not covered by public transportation with an alternative to taxi companies. 
Preliminary studies show that, indeed, ride-hailing services may help improve urban mobility, but they may also increase road traffic, take people off public transport, and generate more pollution \cite{Ngo2021-ot}. 

Moreover, these services may generate segregation because they are based on predictions that discriminate against minorities \cite{Yan2020-kd, Yan2019-aq, Yan_undated-iy}: the models may overfit to strong biases in the source data related to socioeconomic conditions and demographics (see more details in Box 1). 
For example, low ridership in poor neighborhoods is not necessarily (or even typically) an indication of low demand \cite{ge2016racial}.
In general, existing studies are sporadic (due to the algorithms often being regarded as trade secret and algorithmic systems being difficult to audit) and yield inconclusive or contradictory results. 
Further investigation is needed to shed light on these aspects. 

In recent years, there has been a widespread diffusion of GPS navigation apps such as TomTom, Google Maps, and Waze, which use routing algorithms, heuristics and AI to suggest the best path to reach a user’s desired destination. 
Although undoubtedly useful, particularly when exploring an unfamiliar city, navigation apps may also have an adverse impact on the urban environment (see Figure \ref{fig:fig4}c). 

Studies on eco-routing have produced mixed results: green navigation apps can reduce CO2 emissions \cite{Arora2021-xc} but increase exposure to nitrogen oxides \cite{Perez-Prada2017-ps}, and the fastest route suggested by navigation apps may not always optimize fuel consumption \cite{Mehrvarz2020-th}. 
The most recent research \cite{Cornacchia2022-ni} shows that as more vehicles in a city follow the route suggestions of navigation apps, the amount of CO2 emissions in the urban environment increases.
These studies, although still in their early stages, highlight the need for further research on the impact of AI recommendations on route choices and their effects on the urban space and welfare. 
By understanding the interplay between these factors, it may be possible to develop routing algorithms that optimize both fuel efficiency and sustainability while minimizing negative impacts for the environment and society.

\section*{Further Considerations}

Recent computational advances have enabled significant progress in the field, and we anticipate that ongoing research will continue to evolve in tandem with advances in computing. Advancements in data collection methods, such as wearable sensors, smartphones, and connected vehicles, could shed light on poorly understood aspects of mobility. Additionally, emerging techniques for data storage and analysis, such as distributed databases and federated learning, could be instrumental in fusing large amounts of data while maintaining privacy and data quality. Finally, the integration of novel analytical tools and modeling techniques, such as the merging of mechanistic and black-box mobility models, can help uncover new aspects of behavior. These insights could play a crucial role in shaping our society in the years to come. By making progress on the research directions outlined in this Perspective, we may help addressing some of the most pressing challenges that our society faces today, including public health crises, climate change, and socio-economic inequalities. 

\paragraph{Facing public health crises}

Mobility profoundly impacts public health, and mobility data can play a crucial role for understanding the spatial policies that affect it. The COVID-19 pandemic has emphasized this relationship, as many countries have implemented Non-Pharmaceutical Interventions (NPIs) such as lockdowns, stay-at-home orders, and travel restrictions to control the spread of the virus \cite{lai2020effect, haushofer2020interventions, gao2020mobile, chinazzi2020effect, gatto2020spread, jia2020population, tian2020investigation, lucchini2021living, ross2021household, pappalardo2023dataset}. However, traditional NPIs like indiscriminate closure of public and commercial activities may not be the most effective approach. Instead,  an AI model that incorporates diverse data, such as health, demographics, income, mobility, and spatial features, can identify more targeted and effective health policies and NPIs. For instance, if movements to a place are predominantly driven by a specific point of interest, such as a university, limiting access to that location could control the spread of the virus. An AI model could even predict how incoming mobility would change if access to that point was restricted, and give insights into potential outcomes.

\paragraph{Mitigating climate change} 

The relationship between human mobility and its impact on the environment is complex and multi-faceted. As approximately 11.9\% of greenhouse gas (GHG) emissions are from road transport, with 60\% of these emissions resulting from passenger travel, the United Nations (UN) has called for action to reduce the negative environmental impact of cities, with a particular focus on air quality~\cite{ritchie2020sector, assembly2015sustainable}.

One open question in this regard is how to effectively exploit known mobility patterns to define and implement effective policies for reducing emissions.
Studies have shown that a small subset of vehicles, referred to as gross polluters, emit significantly more pollutants compared to other vehicles~\cite{bohm2022gross, nyhan2016predicting}. Encouraging these gross polluters to adopt more sustainable forms of mobility, such as electric vehicles or remote work, is a more effective way to reduce CO2 emissions compared to targeting vehicles randomly~\cite{bohm2022gross}. These findings are promising; however, the relationship between mobility patterns and emissions and sustainability has only been explored on the surface.

Another avenue for further investigation is the development of next-generation routing strategies that consider the collective environmental impact of route choices. The proliferation of mobile navigation apps has further complicated the relationship between mobility and emissions, as the environmental impact of the routes suggested to users can vary considerably from one app to another~\cite{Cornacchia2022-ni, macfarlane2019when, Arora2021-xc}. Designing routing algorithms that reduce the overall environmental impact requires a deeper understanding of human mobility, particularly cognitive factors, to encourage people to follow routes that may be less optimal at the individual level but that collectively help mitigate environmental impact. 
The relationship between multimodality and the environmental impact of mobility, and the potential role of AI in this relationship, remains an area of ongoing research and investigation.

\paragraph{Reducing inequalities}
Achieving sustainable cities that promote social cohesion and where all individuals have equitable access to resources is one of the priorities set by the UN development goals \cite{lu2015policy}. 
Global cities, however, still face deep social inequalities and segregation with respect to aspects such as income, education, and health \cite{glaeser2009inequality}. 
The research directions highlighted in this Perspective could help address this gap, by providing understanding of inequality and segregation driven by mobility patterns.
Segregation and unequal access to opportunities, for example, can be direct consequences of inequitable transport and accessibility. 
Car-centric cities, in fact, favor residential segregation \cite{sanchez2004inequitable}, and poor accessibility to jobs for low-income groups discourages residential moves, thus perpetuating segregation \cite{smith2021sustainable}.
Multimodal transportation promises to help reducing inequalities by offering a range options for getting to locations that are out of reach without a car \cite{carpio2021multimodal} and encouraging social mixing.
The actual effect played by multimodal transportation towards reducing social mixing and inequalities, however, has not been entirely understood, and studies that leverage real-world data and computational models will be key to fill this gap. 
The field of human mobility science can drive novel understanding of social segregation and inequalities and help develop novel solutions to mitigate it by planning for sustainable and inclusive infrastructure. 

\paragraph*{Acknowledgements}
{\footnotesize Luca Pappalardo wishes to thank Giuliano Cornacchia and Giovanni Mauro for their valuable suggestions and Daniele Fadda for his precious support in visualizations.
Luca Pappalardo has been supported by EU projects: 1) H2020 HumaneAI-net G.A. 952026; 2) H2020 SoBigData++ G.A. \#871042; 3) NextGenerationEU - National Recovery and Resilience Plan (Piano Nazionale di Ripresa e Resilienza, PNRR), project “SoBigData.it - Strengthening the Italian RI for Social Mining and Big Data Analytics”, prot. IR0000013, avviso n. 3264 on 28/12/2021.}


\end{document}